\documentclass[12pt]{article}
\usepackage{epsfig,amsmath,amssymb,graphics,color,calc} 
\usepackage{cite}

\usepackage{amsmath} 
\usepackage{amsthm}
\usepackage{amssymb}
\usepackage{amsopn}
\newcommand{\sezione}[2]{ 
\refstepcounter{section}\label{#2} 
\setcounter{equation}{0} 
\setcounter{figure}{0} 
\setcounter{subsection}{0} 
\addcontentsline{toc}{section} 
      {\normalsize\textbf{\thesection.\ #1}} 
\bigskip\bigskip\noindent 
\normalsize\textbf{\thesection.\ #1}\nopagebreak\smallskip\nopagebreak} 
\def\thesection{{\normalsize\arabic{section}}} 
\newcommand{\subsec}[2]{ 
\refstepcounter{subsection}\label{#2} 
\addcontentsline{toc}{subsection} 
      {\normalsize\normalfont\textit{\thesubsection.\ #1}} 
\medskip\medskip\noindent 
\normalsize\normalfont
\textit{\thesubsection. \ #1}\nopagebreak\smallskip\nopagebreak} 
\def\thesubsection{{\normalsize
{\arabic{section}.\arabic{subsection}}}} 
\newcounter{appendice}

\newtheorem{teo}{Theorem}[section]	\newtheorem{pro}[teo]{Proposition}
\newtheorem{defi}[teo]{Definition}	\newtheorem{lem}[teo]{Lemma}
\newtheorem{cor}[teo]{Corollary}	\newtheorem{rem}[teo]{Remark}
\newtheorem{con}[teo]{Condition}

\newcommand{\bteo}[1]{\begin{teo}\label{#1}}
\newcommand{\bpro}[1]{\begin{pro}\label{#1}}
\newcommand{\bdefi}[1]{\begin{defi}\label{#1}}
\newcommand{\blem}[1]{\begin{lem}\label{#1}}
\newcommand{\bcor}[1]{\begin{cor}\label{#1}}
\newcommand{\brem}[1]{\begin{rem}\label{#1}}
\newcommand{\bcon}[1]{\begin{con}\label{#1}}

\newcommand{\eteo}{\end{teo}}	\newcommand{\epro}{\end{pro}}
\newcommand{\edefi}{\end{defi}}	\newcommand{\elem}{\end{lem}}
\newcommand{\ecor}{\end{cor}}	\newcommand{\erem}{\end{rem}}
\newcommand{\econ}{\end{con}}

\renewcommand{\eqref}[1]{(\ref{#1})}

\newcommand{\be}[1]{\begin{equation}\label{#1}}
\newcommand{\bea}[1]{\begin{eqnarray}\label{#1}}
\newcommand{\besn}{\begin{equation*}}
\newcommand{\beasn}{\begin{eqnarray*}}

\renewcommand{\lg}{\left\{}		\newcommand{\rg}{\right\}}

\newcommand{\su}{\subset}	\newcommand{\ssu}{\subset\subset}
\newcommand{\sm}{\setminus}	\newcommand{\es}{\emptyset}

\newcommand{\diam}{\mathop{\rm diam}\nolimits}
\newcommand{\dis}{\mathop{\rm d}\nolimits}
\newcommand{\supp}{\mathop{\rm supp}\nolimits}

	\newcommand{\id}{{1 \mskip -5mu {\rm I}}}
 
\newcommand{\noi}{\noindent}

\newcommand{\ul}[1]{\underline{#1}}

		\renewcommand{\d}{\delta}	
\newcommand{\e}{\varepsilon}	 
		
	\renewcommand{\l}{\lambda}
\newcommand{\m}{\mu}

\newcommand{\s}{\sigma}		
\renewcommand{\t}{\tau}

		\renewcommand{\L}{\Lambda}

\newcommand{\cC}{\mathcal C}	\newcommand{\cD}{\mathcal D} 
	\newcommand{\cF}{\mathcal F} 
	 
\newcommand{\cI}{\mathcal I}	 
	 
	\newcommand{\cN}{\mathcal N}

\newcommand{\cS}{\mathcal S}	\newcommand{\cT}{\mathcal T}

\newcommand{\bE}{\mathbb E}	\newcommand{\bF}{\mathbb F}

	\newcommand{\bL}{\mathbb L} 
	\newcommand{\bN}{\mathbb N} 
	 
	\newcommand{\bR}{\mathbb R} 
	\newcommand{\bT}{\mathbb T} 
	\newcommand{\bV}{\mathbb V} 
	 
	\newcommand{\bZ}{\mathbb Z} 
	\newcommand{\Z}{\mathbb Z}
\newcommand{\newatop}[2]{\genfrac{}{}{0pt}{}{#1}{#2}}

\newcommand{\tree}{\bT}

\renewcommand{\tilde}{\widetilde}
\renewcommand{\dis}{\mathrm d}
\renewcommand{\complement}{\mathrm{c}}
\renewcommand{\cN}{N}

\newenvironment{elencum1}
{\begin{list}{--}{\setlength{\leftmargin}{0.5cm}
\setlength{\itemsep}{-0.1cm}
\setlength{\topsep}{0.cm}}}
{\end{list}}
\newenvironment{elencum2}
{\begin{list}{--}{\setlength{\leftmargin}{0.5cm}
\setlength{\itemsep}{0.1cm}
\setlength{\topsep}{0.3cm}}}
{\end{list}}

\definecolor{light}{gray}{.9}

\begin{document}
\begin{titlepage}
\par\vskip 1cm\vskip 2em

\begin{center}

{\LARGE A combinatorial proof of tree decay of semi--invariants} 
\par
\vskip 2.5em \lineskip .5em
{\large
\begin{tabular}[t]{c}
$\mbox{Lorenzo Bertini}^{1} \phantom{m} \mbox{Emilio N.M.\ Cirillo}^{2}
\phantom{m} \mbox{Enzo Olivieri}^{3}$ 
\\
\end{tabular}
\par
}

\medskip
{\small
\begin{tabular}[t]{ll}
{\bf 1} & {\it 
Dipartimento di Matematica, Universit\`a di Roma La Sapienza}\\
&  Piazzale Aldo Moro 2, 00185 Roma, Italy\\
&  E--mail: {\tt bertini@mat.uniroma1.it}\\
\\
{\bf 2} & {\it
Dipartimento Me.\ Mo.\ Mat., Universit\`a di Roma La Sapienza}\\
&  Via A.\ Scarpa 16, 00161 Roma, Italy\\
&  E--mail: {\tt cirillo@dmmm.uniroma1.it}\\
\\
{\bf 3} & {\it
Dipartimento di Matematica, Universit\`a di Roma Tor Vergata}\\
& Via della Ricerca Scientifica, 00133 Roma, Italy\\ 
& E--mail: {\tt olivieri@mat.uniroma2.it}\\
\end{tabular}
}

\bigskip


\end{center}

\vskip 1 em

\centerline{\bf Abstract} 
\smallskip
We consider finite range  Gibbs fields and 
provide a purely combinatorial proof of the exponential 
tree decay of semi--invariants, supposing that the
logarithm of the partition function can be expressed as a sum 
of suitable local functions of the boundary conditions. 
This hypothesis holds for completely analytical Gibbs fields; 
in this context the tree decay of semi--invariants has been 
proven via analyticity arguments.
However the combinatorial proof given here
can be applied also to the more complicated case of 
disordered systems in the so called Griffiths' phase when analyticity
arguments fail.

\vskip 0.8 em

\vfill
\noindent    
{\bf MSC2000:} 82B20, 60G60  

\vskip 0.8 em
\noindent
{\bf Keywords:}\ 
Gibbs Fields, Semi--invariants, Cluster expansion, Disordered systems

\bigskip\bigskip
\footnoterule
\vskip 1.0em
{\small 
\noindent
L.B.\ and E.O.\ are partially supported by Cofinanziamento MURST.
E.C.\ thanks the Istituto Nazionale di Alta Matematica -- GNFM for 
financial support. 
\vskip 1.0em
\noindent
}
\end{titlepage}
\vfill\eject

\sezione{Introduction}{s:i}
\par\noindent 
In this note we present a purely combinatorial proof of the tree decay of
semi--invariants, also called truncated correlations, Ursell functions, or
cumulants, 
for a finite range Gibbsian field under the condition that the
logarithm of the partition function can be expressed as the
sum of suitable local functions of the boundary condition.

Let $Z_\L (\t)$ be the partition function in the finite volume 
$\L\subset\bZ^d$ 
with boundary condition $\t$ outside $\L$; we assume that
\be{basic} 
\log Z_\L (\t) \,=\, \sum _{X\subset\bZ^d:\,X\cap \L \neq \emptyset} 
  \phi_{X,\L}(\t)
\end{equation}
where the ``effective potentials" $\phi_{X,\L}$ are such that:
\begin{itemize}
\item[(i)]
\label{i:int1}
given $X\subset\bZ^d$, the functions $\phi_{X,\L}$ are constant w.r.t.\ 
$\L$ for the $\L$'s with a given intersection with $X$;
\item[(ii)]
\label{i:int2}
have a suitable decay property with the size of $X$, uniformly in $\L$.
\end{itemize}

The expression (\ref{basic}) can be obtained via cluster expansion in 
the weak coupling
(high temperature and/or small activity) region but it holds in more general
situations. It can also be obtained in the framework of  Dobrushin--Shlosman
complete analyticity as well as in the framework of the so--called  
{\it scale--adapted cluster expansion}, 
see \cite{[BCOabs]}, provided the volume
$\L$ is a disjoint union of cubes whose side length equals the scale 
of the expansion.
We refer to \cite{[BCOabs]} 
for a more exhaustive discussion; here we only say that 
scale--adapted cluster expansions have been introduced in \cite{[O],[OP]}
in order to perturbatively treat the whole uniqueness region of 
lattice spin systems, arbitrarily close to the coexistence line. 
Moreover, as we shall see in \cite{[BCOran]}, a variant of (\ref{basic}) 
holds  in the
context of disordered lattice systems, also in the delicate
situation of Griffiths' singularity that makes necessary the use of a 
{\it graded cluster expansion}, see \cite{[BCOabs]}.

In the framework of the renormalization group maps one often 
encounters an expression like (\ref{basic}) for the renormalized 
partition function. In that case the family 
$\{\phi_{X,\Lambda},\,X\subset\bZ^d\}$
represents the ``finite--volume renormalized potential".
Both in the case of disordered systems and  
of renormalization group maps the decay
properties of $\phi_{X,\Lambda}$ are weaker than the 
corresponding ones of the 
case of weakly coupled short range Gibbs fields

The tree decay of semi--invariants is often deduced from analyticity
properties of the pressure, 
see \cite{[DS3],[DuIaSo1],[DuIaSo2],[DuSo],Ru};
however, there are physically interesting situations in 
which these analyticity properties do not hold but nevertheless we expect 
the exponential decay of semi--invariants. The main example is given 
by the already quoted case of a disordered lattice spin system, like a 
spin glass or a ferromagnetic system subject to a random field, in 
presence of the so--called ``Griffiths' singularity".
Consider, for example, a random coupling Ising spin system in $\bZ^d$ 
described by the formal Hamiltonian: 
\be{introham} 
H(\s) = \sum _{x,y:|x-y|=1} J_{x,y}\s_x\s_y - h \sum _x \s_x
\end{equation}
where $\s_x \in \{-1,+1\}$,  $h\in \bR$ is fixed and  
$J_{x,y}$ are i.i.d.\ Gaussian random 
variables with mean zero and variance one.
At high temperature we expect an exponential tree decay of semi--invariants 
with a deterministic rate (this has actually been proved long time ago 
in \cite{[FI]}) but we do not expect analyticity of thermodynamic 
functions. This behavior is a consequence of the fact that, even though 
in average the system is weakly coupled nonetheless, with a positive 
probability, arbitrarily large regions with strong ferromagnetic couplings 
can appear inducing, locally, long--range order, as a consequence of the 
unboundedness of the random couplings.

The starting point of our combinatorial computation can be illustrated 
in the simple case of the semi--invariant of order two namely, the covariance
between two local functions, see \cite{BCC}.
For instance,
consider a lattice spin system with finite state space and finite
range interaction, say $r\in[0,\infty)$, 
whose Hamiltonian, in a finite box $\L$ for a
configuration $\s_\L$ in
$\L$ and a boundary condition $\t_{\L^c}$ is denoted by
$H_\L(\s_\L\t_{\L^c})$. More detailed and precise definitions will be 
given later on; here we only say that 
$H_\L(\s_\L\t_{\L^c})$ contains the self--interaction of $\s_\L$ in $\L$ 
and the mutual interaction between $\s_\L$ and $\t_{\L^c}$.  
The Gibbs measure is 
$\m^\t_\L(\s_\L)=\exp\big\{H_\L(\s_\L\t_{\L^c})\big\}/Z_\L(\t_{\L^c})$
where $Z_\L(\t_{\L^c})=\sum_{\s_\L}\exp\big\{H_\L(\s_\L\t_{\L^c})\big\}$. 
Notice that we have included 
the inverse temperature in $H_\L$ and changed the usual 
convention on the sign in the exponent.

Let $f,g$ be local functions with supports $\L_f,\L_g\subset \L$ such that
${\rm dist} (\L_f,\L_g)> r$. We may write
\begin{equation}
\label{BCC}
\begin{array}{rcl}
\m_\L^\t(f;g)
&=&\m_\L^\t(fg)-\m_\L^\t(f)\,\m_\L^\t(g)\vphantom{\Bigg\{}\\
&=& 
 {\displaystyle
 \sum_{\s_{\L_f},\s_{\L_g}} f(\s_{\L_f})g( \s_{\L_g})\,
 e^{H_{\L_f}(\s_{\L_f}\t_{\L^c})}e^{H_{\L_g}(\s_{\L_g}\t_{\L^c})}}\\
&& 
 {\displaystyle
  \;\;\;\;\times
  \Big(
  \frac{Z_{\L \setminus (\L_f\cup\L_g)}(\s_{\L_f} \s_{\L_g}\t_{\L^c})}
       {Z_{\L}(\t_{\L^c})}
  - 
  \frac{Z_{\L\setminus\L_f}(\s_{\L_f}\t_{\L^c})
         Z_{\L \setminus  \L_g }( \s_{\L_g}\t_{\L^c})}
       {Z_{\L}^2(\t_{\L^c})}\Big)}\\
\end{array}
\end{equation}
It is clear that the exponential decay of $\m_\L^\t(f;g)$ with 
${\rm dist} (\L_f,\L_g)$ easily follows from the analogous property 
of the quantity 
\be{condiz}
\sup_{\s_{\L_f},\s_{\L_g},\t_{\L^c}}
\bigg| 
\frac{Z_{\L \setminus (\L_f\cup\L_g)}(\s_{\L_f} 
      \s_{\L_g}\t_{\L^c})Z_{\L}(\t_{\L^c})}
     {Z_{\L\setminus\L_f}(\s_{\L_f}\t_{\L^c}) 
      Z_{\L\setminus\L_g }(\s_{\L_g}\t_{\L^c})}-1
\bigg|
\end{equation}
This, in turn, is easily seen to follow from (\ref{basic}) and suitable decay 
properties of 
$\phi_{X,\L}$, see (\ref{tm2}) below.
Indeed by plugging (\ref{basic}) into (\ref{condiz}) and using 
(i) above we easily see that, in the resulting expression,
$\phi_{X,\Lambda}$ cancels out unless $X$ intersects both $\L_f$ and $\L_g$.

The case of a generic semi--invariant of order $n$ is much more subtle and
some more efforts are required to disclose the cancellation mechanism.
The crucial point in our proof is the combinatorial result
in Lemma~\ref{t:srpf} which generalizes (\ref{BCC}) and  
expresses the semi--invariant in terms of ratios of partition 
functions. 

The paper is organized as follows. In Section~\ref{s:nr}
we give the notation and a 
theorem stating our main result, with some comments and exempla. 
The proof of the theorem is finally given in Section~\ref{s:pr}.

\sezione{Notation and result}{s:nr} 
\par\noindent 
In this Section we recall the general framework of Gibbs states 
for lattice systems, state our main results, and discuss some possible
applications. 

\subsec{The lattice}{s:lat} 
\par\noindent
For $a,b\in\bR$ we set $a\wedge b:=\min\{a,b\}$ and 
$a\vee b:=\max\{a,b\}$. 
For $x=(x_1,\cdots,x_d)\in\bR^d$ we set $|x|:=\sup_{k=i,\cdots,d}
|x_i|$.  The spatial structure is modeled by the $d$--dimensional
cubic lattice $\bL:=\Z^d$. We shall denote by $x,y,\cdots$ the points
in $\bL$, called {\em sites}, and by $\L,V,X,\dots$ the subsets of $\bL$.
We use $\L^\complement:=\bL\setminus \L$ to denote the complement of
$\L$.  For $\L$ a finite subset of $\bL$, we use $\L\subset\subset\bL$
to indicate that $\L$ is finite, $|\L|$ denotes the cardinality of
$\L$.  We consider $\bL$ endowed with the distance
$\dis(x,y)=|x-y|$. As usual for $X,Y\subset\bL$ we set
$\dis(X,Y):=\inf\{\dis(x,y),\; x\in X,\; y\in Y\}$,
$\diam(X):=\sup\{\dis(x,x'),\; x,x'\in X\}$.  

For $x\in\bL$ and $m$ a positive integer we let 
$Q_m(x):=\{y\in\bL:\,x_i\le y_i\le x_i+(m-1)\,,\: i=1,\dots,d\}$
be the cube of side $m$ with $x$ the site with smallest coordinates.
We denote by $\bF :=\{ X\ssu \bL\}$ the collection of all 
finite subsets of $\bL$. Let $L$ be a positive integer, we denote by
$\bF_L$ the collection of sets in $\bF$ which can be written as
the disjoint union of cubes of side $L$, more precisely $X\in\bF_L$ iff
there exist $x_1,\dots,x_k\in\bL$
such that $X=\bigcup_{h=1}^k Q_L(Lx_h)$.

Let $\bE:=\big\{\{x,y\},\,x,y\in\bL:\,\dis(x,y)=1\big\}$ be the collection
of {\em edges\,} in $\bL$.  Note that, according to our definitions, the
edges can be also diagonal.  We say that two edges
$e,e'\in\bE$ are connected iff $e\cap e'\not=\emptyset$.  
A subset $(V,E)\su (\bL,\bE)$ is said to be connected iff for each
pair $x,y\in V$, $x\neq y$,
there exists in $E$ a path of connected edges joining
them. For $X\ssu\bL$ we then set 
\be{treedec0} 
\tree(X):=\inf\lg |E| \,, \ (V,E) \su (\bL,\bE)\:
\textrm{connected}: \: V \supset X \rg
\end{equation}
and remark that the infimum is attained (not
necessary uniquely) for a graph $(V_X,E_X)\subset(\bL,\bE)$ which
is a tree, i.e.\ a connected and loop--free graph.
We agree that $\tree(X)=0$ if $|X|=1$ and note that for $x,y\in\bL$ we have 
$\tree(\{x,y\})=\dis(x,y)$.

\subsec{The configuration space}{s:conf} 
\par\noindent
The {\em single spin space\,} is given by a finite set
$\cS_0\subset\bR$ which we consider endowed with its discrete
$\s$--algebra $\cF_0$. 
The configuration space in $\Lambda\subset\bL$ is
$\cS_\Lambda:=\cS_0^\L$ equipped with the product
$\s$--algebra $\cF_\L=\cF_0^\L$; we
denote $\cS_0^{\bL}$ and $\cF_0^\bL$ simply by $\cS$ and $\cF$.
Elements of $\cS$, called {\it configurations\,}, are denoted by
$\sigma,\tau,\dots$. In other words a configuration $\s\in\cS$
is a function $\s:\bL\to\cS_0$; for $\L\su\bL$ we denote by
$\sigma_{\Lambda}$ the restriction of $\sigma$ to $\Lambda$.
Let $\L_1,\L_2\subset\bL$ be disjoint subsets of $\bL$; 
if $\sigma_i\in\cS_{\L_i}$, $i=1,2$, we denote by 
$\s_1\s_2$ the configuration in $\cS_{\L_1\cup\L_2}$ given by
$\s_1\s_2(x):=\sum_{i=1}^2\id_{\{x\in\L_i\}}\s_i(x)$ for any
$x\in\Lambda_1\cup\Lambda_2$.

A measurable function $f:\cS\rightarrow\bR$ is called a {\em local
function\,} iff there exists $\Lambda\in\bF$ such that
$f\in\cF_{\Lambda}$, namely $f$ is $\cF_{\Lambda}$--measurable for
some $\Lambda\in\bF$.
For $f$ a local function we shall denote by $\supp(f)$, the so called
{\em support\,} of $f$, the smallest $\Lambda\subset\subset\bL$ such that
$f\in\cF_{\Lambda}$. If $f\in\cF_\Lambda$ we shall sometimes abuse the
notation by writing $f(\sigma_\Lambda)$ instead of $f(\sigma)$. 
For $f\in\cF$ we let $\|f\|_\infty:=\sup_{\s\in\cS}|f(\s)|$ be the sup
norm of $f$.

\subsec{The Gibbs state}{s:gibbs} 
\par\noindent
A {\em potential\,} $U$ is a collection of local functions
$U_X:\cS\rightarrow\bR$, $\cF_X$--measurable,
labeled by finite subsets of $\bL$,
namely $U:=\{U_X\in\cF_X,\,X\in\bF\}$.
We shall consider only {\em finite range\,} potential
namely, potentials $U$ for which there exists an integer $r$, called
{\em range\,} such that $U_X=0$ if $\diam(X)>r$.
We remark that we do not require the
potential $U$ to be translationally invariant.

For $\Lambda\subset\subset\bL$ and 
$\s\in\cS$ we define the {\em Hamiltonian} as
\be{ham}
H_{\L}(\sigma):=\sum_{X\cap\L\neq\es} U_X(\s)
\end{equation}
In this paper we shall consider only finite volume Gibbs measures
defined as follows: let $\tau\in\cS$, the finite volume Gibbs measure
$\mu_{\Lambda}^{\tau}$, with boundary condition $\tau$, is the
probability measure on $\cS_\Lambda$ given by
\be{ls-sone} 
\mu_\Lambda^\t (\sigma):=\frac{1}{Z_{\L}(\t)}
                         \:e^{H_\Lambda(\sigma\tau_{\Lambda^\complement})}
\end{equation}
where $\sigma\in\cS_\Lambda$ and $Z_\Lambda(\tau)$, called the
{\it partition function}, is the normalization constant given by 
\be{Z}
Z_{\L}(\t):=\sum_{\sigma\in\cS_\Lambda}  
e^{H_{\L}(\s \tau_{\L^\complement})}
\end{equation}
we remark that, 
since the potential $U$ has range $r$, we have
$Z_\L\in \cF_{\{ x\in\L^\complement\,:\, \dis(x,\L) \le r \}}$.  

For $V\su\L\ssu\bL$ we shall denote by $\mu_{\L,V}^\tau$ the
projection (marginal) of $\mu_{\L}^\tau$ to $\cS_V$ namely, the
probability measure on $\cS_V$ given by $\mu_{\L,V}^\tau(A) =
\mu_{\L}^\tau(A)$, $A\in\cF_V$. 

\subsec{Semi--invariants}{s:sin} 
\par\noindent
Let $\L\in\bF$, $n\ge 2$ an integer, $f_i$, with $i=1,\dots,n$, 
local functions with $\Lambda_i:=\supp(f_i)\subset\Lambda$,
$t_i\in\bR$, with $i=1,\dots,n$, and $\tau\in\cS$; we define 
\be{pZ}
Z_\Lambda\big(\tau;t_1,\dots,t_n\big)
:= \mu_\Lambda^\tau \bigg( \exp\Big\{\sum_{i=1}^n t_i
f_i\Big\} \bigg)
\end{equation}
The semi--invariant of $f_1,\dots,f_n$ w.r.t.\ the finite volume Gibbs
measure $\mu_\Lambda^\tau$ is then defined by
\begin{equation}
\label{2.1}
\mu_\Lambda^\tau \big( f_1;\cdots;f_n\big) :=
\frac{\partial^n
      \log Z_\Lambda\big(\tau;t_1,\dots,t_n)}
{\partial t_1\cdots \partial t_n}
\Bigg|_{t_1=\cdots=t_n=0}
\end{equation}
note that for $n=2$ we have 
$\mu_\Lambda^\tau \big( f_1;f_2\big)
=\mu_\Lambda^\tau \big( f_1\, f_2\big) - \mu_\Lambda^\tau \big( f_1)
\mu_\Lambda^\tau \big(f_2)$ namely, the covariance between $f_1$ and $f_2$.

It is possible to express the semi--invariant in terms of the moments of
$f_1,\dots, f_n$. For notation compactness let us set
$\cN:=\{1,\dots,n\}$ and denote by $\cD_N^\ell$ the collection of the
partitions of $\cN$ into $\ell$ atoms namely,
\begin{equation}
\label{2.2}
\cD_N^\ell:=\Big\{\ul{D}\equiv\big\{D_1,\dots,D_\ell\big\}:
D_i \subset \cN ,\: 
D_i\neq\emptyset,\: 
D_i\cap D_j =\emptyset\textrm{ for }
i\neq j,\: \bigcup_{i=1}^\ell D_i =\cN \Big\}
\end{equation}
We then have, see e.g.\ \cite[II, \S 12.8]{Sh}
\begin{equation}
\label{sm}
\mu_\Lambda^\tau \big( f_1;\cdots;f_n\big) =
\sum_{\ell=1}^n (-1)^{\ell -1} (\ell-1)! 
\sum_{\ul{D}\in\cD_N^\ell}\,
\prod_{k=1}^\ell\,
 \mu_\Lambda^\tau\Big(\prod_{i\in D_k}f_i\Big)
\end{equation}

\subsec{Tree decay of semi--invariants}{s:mr} 
\par\noindent
We may now state our main result. Let $f_1,\dots,f_n$ be local
functions, $n\ge 2$. Given a positive integer $L$, by enlarging
$\Lambda_i:=\supp(f_i)$, we may (and do) assume that
$\Lambda_i\in\bF_L$; we shall further assume that for $i\neq
j\in\cN$ we have $\dis(\Lambda_i,\Lambda_j)> r$.
We stress that the supports $\Lambda_i$ can be arbitrarily large, 
possibly diverging with $\Lambda$. 

Let us denote by $(\bV_{\cN},\bE_{\cN})$ the graph
obtained from $(\bL,\bE)$ by contracting each $\Lambda_i$, $i\in \cN$, to a
single point, in other words we define
$\bV_{\cN}:=\{x\,:\:x\in\bL\sm\bigcup_{i=1}^n\Lambda_i\}
\cup\bigcup_{i=1}^n\{\Lambda_i\}$, 
$\bE_{\cN}:=\{ \{v,v'\},\,v,v'\in\bV_{\cN}\,:\ \dis(v,v')=1\}$, and 
\begin{equation}
\label{tif}
\cT\big(f_1;\dots;f_n\big):=
\inf\Big\{ |E| \,, \ (V,E) \su (\bV_N,\bE_N)\:
\textrm{connected}: \: V \supset \bigcup_{i=1}^n \{\Lambda_i\} \Big\}
\end{equation}

\begin{teo}
\label{t:princ}
Let $L\in\bN$,
assume that for each $\Lambda\in\bF_L$ and $\tau\in\cS$
we have the expansion
\begin{equation}
\label{tm1} 
\log Z_{\Lambda}(\tau)= 
\sum_{X\cap\Lambda\not=\emptyset} \phi_{X,\Lambda}(\tau)
\end{equation}
for some local functions
$\phi_{X,\Lambda}\in\cF_{\Lambda^\complement}$,
$X\in\bF$, such that given
$\Lambda,\Lambda'\subset\subset\bL$, we have
that $X\cap\Lambda=X\cap\Lambda'$ implies
$\phi_{X,\Lambda}=\phi_{X,\Lambda'}$.
If there exist reals $a,b\ge 0$ 
and $C<\infty$ such that for any $\L\in\bF_L$
\begin{equation}
\label{tm2}
\sup_{x\in\bL}\,
 \sum_{X\ni x}\exp\{a\tree(X)+ b\diam(X)\}
  \, \left\|\phi_{X,\Lambda}\right\|_\infty
\le C
\end{equation}
then for each $n\ge 2$ there exists a real
$K_n=K_n(C;|\L_1|,\dots,|\L_n|)$ such that 
\begin{equation}
\label{tm80}
\big|\mu_\Lambda^\tau(f_1;\dots;f_n)\big|
\le K_n\:
\exp\Big\{-\Big[a+\frac{b}{n-1}\Big] \,\cT(f_1;\dots;f_n)\Big\}\:
\prod_{i=1}^n\mu_\Lambda^\tau(|f_i|)
\end{equation}
for any $\Lambda\in\bF_L$ and $\tau\in\cS$.
\par\noindent
Furthermore, if \eqref{tm2} is satisfied with $a>0$
and $\Lambda_1,\dots,\Lambda_n$ are such that for some $\d\in(0,1)$ 
\begin{equation}
\label{tm6}
\lambda:=
\sup_{i\in\cN}
\sum_{j\not=i}
 \big( |\Lambda_i|\wedge|\Lambda_j| \big)\,
  \exp\Big\{-\frac{1}{2} \,a\, \d \, \dis(\Lambda_i,\Lambda_j) \Big\}
\le \frac{1}{6\,e\,(1+18\,C\,e)}
\end{equation}
then 
\begin{equation}
\label{tm8}
\big|\mu_\Lambda^\tau(f_1;\dots;f_n)\big|
\le
\exp\Big\{-a\,(1-\delta)\,\cT(f_1;\dots;f_n)\Big\}\:
\prod_{i=1}^n\mu_\Lambda^\tau(|f_i|)
\end{equation}
for any $\Lambda\in\bF_L$, $\tau\in\cS$, and $n\ge 2$.
\end{teo}
Note that the hypotheses (\ref{tm1}) and (\ref{tm2}) with $a+b>0$ imply
\cite[Condition IVa]{[DS3]} which is one of the Dobrushin--Shlosman
complete analyticity conditions. Indeed by setting 
\begin{equation}
\label{eqrp}
g(x,\Lambda,\tau):=\sum_{X\ni x}\frac{1}{|X|}\,\phi_{X,\Lambda}(\tau)
\end{equation}
for all $\tau\in\cS$, $\Lambda\subset\subset\bL$, and $x\in\Lambda$, we 
have that (i) e (ii) of \cite[Condition IVa]{[DS3]} hold.

\begin{rem}
\label{3}
\normalfont{
Instead of (\ref{tm1}) we can assume an expansion of the form 
\be{tm1'}
\log Z_\Lambda(\tau)=\sum_{X\cap\Lambda\not=\emptyset}
\big[\psi_{X,\Lambda}(\tau) +\phi_{X,\Lambda}(\tau)\big]
\end{equation}
where $\phi_{X,\Lambda}$ satisfies the bound \eqref{tm2} whereas
$\psi_{X,\Lambda}$ satisfy the same measurability condition namely,
that $\psi_{X,\Lambda}\in\cF_{\Lambda^\complement}$ and
$X\cap\Lambda=X\cap\Lambda'$ implies
$\psi_{X,\Lambda}=\psi_{X,\Lambda'}$, $X\in\bF$, and
$\psi_{X,\Lambda}$ are of finite range, i.e.\ for some integer $\bar r$ we
have $\psi_{X,\Lambda}=0$ for $\diam(X)>\bar r$.  Then the thesis of
Theorem~\ref{t:princ} still holds provided $\dis(\Lambda_i,\Lambda_j)>\bar r$,
$i\neq j\in \cN$.
Note that no bound on the norm of the family 
$\{\psi_{X,\Lambda},\,X\in\bF\}$ is required.
}
\end{rem}

\medskip
\noi{\it Addenda:}
\begin{elencum1}
\item
One may wonder how we can bound the semi--invariant of $n$ functions
in terms of their $L^1$ (rather than $L^n$) norm.  This is possible
because $f_i$ have disjoint supports.
\item
By the methods in
\cite{[O],[OP]}, it is possible to prove the following converse to
Theorem~\ref{t:princ}. If the bound \eqref{tm80} holds for $n=2$ then
there are an integer $L'>0$ and a real $a'>0$  such that \eqref{tm1}
and \eqref{tm2} hold for any $\L\in\bF_{L'}$. 
\item
If there exists a unique infinite volume Gibbs state $\mu$, as it is
typically the case under conditions implying 
the validity of \eqref{tm1}--\eqref{tm2},
then the bounds \eqref{tm80} and \eqref{tm8} holds also for $\mu$.
\item
If the supports $\Lambda_i$ are at distance large enough (depending on
$|\Lambda_i|$, $a$ and $C$), then the condition \eqref{tm6} is satisfied.
Note also that {\em one\,} of the functions $f_i$ might have
arbitrarily large support. 
\item
In \cite[Corollary II.12.8]{[Sim]} it is shown how,
in a general setting, it is possible to deduce some decay of 
semi--invariants from suitable decay properties of covariances.
\end{elencum1}

\subsec{Exempla}{s:ex} 
\par\noindent 
In order to clarify how (\ref{tm1}) and (\ref{tm2}) can be shown to hold 
assuming a convergent cluster expansion,
we discuss the standard Ising model at high temperature;
much more general models can be analyzed along the same lines.  
The single spin configuration space is $\cS_0=\{-1,+1\}$ and 
the potential $U$ is then given by 
$$    
U_X(\sigma):=\left\{
\begin{array}{ll}
J\sigma(x)\sigma(y) & \;\;\;\;\;\textrm{if } X=\{x,y\}\textrm{ and }
                                |x-y|_2=1 \\
0 & \;\;\;\;\;\textrm{otherwise}
\end{array}
\right.
$$
where $J\in\bR$ and $|x|_2$ is the Euclidean norm of $x\in\bZ^d$.
The partition function (\ref{Z}) can be written as
$$
Z_\Lambda(\tau)=
2^{|\Lambda|}
\bigg[
 1+\sum_{n\ge1}\sum_{\newatop{\gamma_1,\dots,\gamma_n\in\Gamma_\Lambda:}
                             {\tilde\gamma_i\cap\tilde\gamma_j=\emptyset
                              \;\;\;1\le i<j\le n}}
               \prod_{k=1}^n\zeta_{\gamma_k}(\tau)
\bigg]
$$
where $\Gamma_\Lambda$ is the set of polymers intersecting $\Lambda$;
a polymer $\gamma\in\Gamma_\Lambda$ is a connected set of bonds:
for some $k\ge1$, $\gamma=\{b_1,\dots,b_k\}$ with 
$b_i=\{x_i,y_i\}$, $|x_i-y_i|_2=1$, $b_i\cap\Lambda\neq\emptyset$.
We have also set $\widetilde\gamma:=\cup_{b\in\gamma}b$ and 
$$
\zeta_\gamma(\tau):=\frac{1}{2^{|\Lambda|}}
                    \sum_{\sigma\in\{-1,+1\}^\Lambda}
                    \prod_{b\in\gamma}
                    \big[
                      e^{U_b(\sigma\tau_{\Lambda^\complement})}-1
                    \big]
$$
note that for each $\gamma\in\Gamma_\Lambda$ we have 
$\zeta_\gamma\in\cF_{\widetilde\gamma\cap\Lambda^\complement}$.
For $|J|$ small enough it is possible to show, see 
e.g.\ \cite[\S 20.4]{GJ} or \cite[\S V.7]{[Sim]}, that 
\begin{equation}
\label{expising}
\log Z_\Lambda=\sum_{n=1}^\infty
               \sum_{\gamma_1,\dots,\gamma_n\in\Gamma_\Lambda}
                 \varphi_T(\gamma_1,\dots,\gamma_n)
                 \prod_{k=1}^n\zeta_{\gamma_k}
\end{equation}
where $\varphi_T$ is a combinatorial factor, see e.g.\ 
\cite[Eq.~(20.2.8)]{GJ} or \cite[Eq.~(V.7.9)]{[Sim]}, vanishing whenever 
$\{\gamma_1,\dots,\gamma_n\}$ can be split into two subsets 
with every polymer of the first one not intersecting any polymer 
of the second one. From (\ref{expising}) we get (\ref{tm1}) with 
$\phi_{X,\Lambda}$ given by
$$
\phi_{X,\Lambda}=\sum_{n=1}^\infty
                 \sum_{\newatop{\gamma_1,\dots,\gamma_n\in\Gamma_\Lambda:}
                               {\bigcup_{i=1}^n\widetilde\gamma_i=X}}
                  \varphi_T(\gamma_1,\dots,\gamma_n)
                 \prod_{k=1}^n\zeta_{\gamma_k}
$$
Finally, by standard estimates, see e.g.\ 
\cite[\S 20.4]{GJ} or \cite[\S V.7]{[Sim]}, we get that the bound (\ref{tm2}) 
holds for some $a>0$. 
\smallskip

Without entering into the details, we discuss here some models to which
Theorem~\ref{t:princ} might be applied on the basis of a
convergent cluster expansion. 

\begin{elencum2}
\item
{\em High temperature / low activity expansions.}\hfill\break 
The convergence of the cluster expansion for any $\L\in\bF$ and the
tree decay of the semi--invariants for $f_i(\s)=\s(x_i)$, $x_i\in\bL$
is a classical topic in equilibrium statistical mechanics, see e.g.\
\cite[\S 20.4]{GJ}, \cite[Theorem V.7.13]{[Sim]},
and \cite{[DuIaSo1],[DuIaSo2],[DuSo]}.
However we are not aware of any reference where the case of 
local functions $f_i$ with arbitrary support is discussed in detail. 

\item{\em Strong Mixing (SM) potentials.}\hfill\break 
The tree decay of the semi--invariants
uniform in the boundary configuration is one, called condition IIc, of
the equivalent conditions of the Dobrushin--Shlosman's
completely analytical interactions \cite{[DS2],[DS3]}. It is
stated in a somewhat different form than the one given here: there is
no restriction on $\dis(\Lambda_i,\Lambda_j)$, but the supports $\Lambda_i$ are
required to have $\diam(\Lambda_i)\le r$.  
We mention that the equivalence of the tree decay of the
semi--invariants with the other conditions is proven, via a very
elegant analytical function argument, under the additional assumption
that the potential $U$ is in the same connected component (among the
interactions satisfying the conditions) of the zero potential, see
\cite[Comment 2.1]{[DS2]}.  

In the original Dobrushin--Shlosman's setting the exponential decay
\eqref{tm80} is supposed to hold for all $\L\in\bF$; however, as
discussed in \cite{[MO2]}, 
there are examples in which it holds only for $\L\in\bF_L$ with
$L$ large enough. This has lead to the so--called restricted
completely analytical (or Strong Mixing) scenario, see
\cite{[MO2],[MOS],[ScS]}, in which one considers only the ``regular''
volumes $\L\in\bF_L$.
The usual argument to get the tree decay of the semi--invariants for
SM potentials is the following.  Consider a rescaled system whose new
single spin variables are the old spin configurations in the blocks
$Q_L(Lx)$, $x\in\bZ^d$; we can then apply Dobrushin--Shlosman's
results \cite{[DS2],[DS3]} 
to this rescaled system and get all their equivalent mixing and
analyticity properties of the Gibbs state for every $\L\in\bF_L$.

Theorem~\ref{t:princ} allows a direct proof of the tree decay of
semi--invariants for SM potentials (without the hypotheses that $U$ is
in the same connected component of the zero potential) according to
the following route. SM potentials satisfy the {\em finite size\,}
condition introduced in \cite{[O],[OP]} which yields a convergent
cluster expansion for which \eqref{tm1} and \eqref{tm2} hold for some
$a>0$ and some integer $L$. As a matter of fact in \cite{[O],[OP]}
it is considered only the case when $\L$ is a torus, but it is not too
difficult, see \cite{[BCO],[BCObat]} 
for some details, to extend it to any $\L\in\bF_L$
and $\tau\in\cS$. Then Theorem~\ref{t:princ} yields the tree decay of
the semi--invariants in the sense given by \eqref{tm8}.

\item{\em Continuous systems. High temperature / low activity
expansions.}\hfill\break 
We have described only lattice models, but it is possible to
extend Theorem~\ref{t:princ} to continuous models. For the infinite
volume state, absolute integrability of the Ursell functions is proven
in \cite[Thm. 4.4.8]{Ru}. 
For a positive pairwise interaction, the convergence of the cluster
expansion uniform in the boundary condition is proven in \cite{Sp},
see also \cite{BCC} for the exponential decay of the covariance
between local functions.

\item{\em Disordered systems in the Griffiths' phase. High temperature /
low activity expansions.}\hfill\break
The convergence of an appropriate multi--scale cluster expansion in such
a situation has been obtained in \cite{[FI]} where the tree decay of the
semi--invariants is proven in  
detail only for $f_i(\s)=\s({x_i})$. We are in a situation like the
one described in Remark~\ref{3} with the additional complication that,
depending on the disorder configuration, the functions $\psi_{X,\L}$
can have arbitrary large supports. One then obtains some probability
estimates on the disorder which lead to a tree decay in a set of full measure.

\item{\em Disordered systems in the Griffiths' phase. Small perturbation
of SM potentials.}\hfill\break
The convergence of an appropriate multi--scale cluster expansion in such
a situation will be proven in \cite{[BCOran]}. We stress that, due to
the presence of arbitrary large regions of strong interaction, the
bound \eqref{tm2} holds with $b>0$ but $a=0$.
We refer to \cite{[BCOabs]} for a more detailed discussion.
\end{elencum2}

\sezione{Proof of the tree decay}{s:pr} 
\par\noindent 
The usual proofs of tree decay of the semi--invariants from the
convergence of the cluster expansion, see e.g.\ \cite[\S 20.4]{GJ} 
or \cite[\S V.7]{[Sim]}, are
based on the expansion of the perturbed partition function \eqref{pZ}
and then in the estimates of the derivatives in \eqref{2.1}. If one is
willing to consider functions $f_i$ with arbitrary supports $\Lambda_i$
there are some difficulties related in the need of cluster expand the
perturbed measure also inside $\Lambda_i$ where the interaction is not
necessary weak. The combinatorial proof we present here is instead
based on the identity \eqref{sm} and will involve \eqref{tm1}
and \eqref{tm2}, which abstract the convergence of a
cluster expansion, only outside the supports $\Lambda_i$ namely, for the
unperturbed system. For simplicity we have required that the
supports of the functions $f_i$ are at a
distance greater than the range of the potential.

\smallskip
Let us start by a purely combinatorial lemma which reduces the
estimate of the semi--invariant to ratios of partition functions.
For $\Lambda\in\bF_L$ and
$I\subset\cN=\{1,\dots,n\}$ we set 
$\Lambda_I:=\bigcup_{i\in I}\Lambda_i\subset\L$ and 
$V_I:=\Lambda\setminus\Lambda_I$; 
note that since we have assumed $\Lambda_i\in F_L$ we have also $V_I\in F_L$.
For $\sigma\in\cS$ let also $R_I=R_I(\sigma)$ be defined by 
$R_{\emptyset}=R_{\{i\}} = 1$ and for $|I|\ge 2$ by
\begin{equation}
\label{2.3}
\log R_I := \sum_{J\subset I} (-1)^{|I|-|J|} \log Z_{V_J}
\end{equation}
note that $R_I\in\cF_{V_I^\complement}$ and 
$V_I^\complement=\Lambda^\complement\cup\Lambda_I$. 
We point out the analogy between definition (\ref{2.3}) and the 
combinatorial set up of Koteck\'y--Preiss \cite[Eq.~(3)]{[KP]}.
We set finally $\varrho_I:=
R_I-1$ and define $\varrho^{(\L)}\in\cF_{\L_{\cN}\cup\L^\complement}$ as 
\begin{equation}
\label{2.3.1}
\varrho^{(\L)}(\sigma):=
\sum_{k\ge 1}\:
 \sum_{\ul{I}\in \cI_k}\:
  \prod_{I\in\ul{I}}\,
   \varrho_I (\sigma)
\end{equation}
in which 
\begin{equation}
\label{2.4}
\cI_k:=\bigg\{ \ul{I} \equiv \big\{I_1,\dots,I_k\big\}\,:\:
I_h\subset\cN\,, \:  h\neq h'\Rightarrow I_h\neq I_{h'}\,,\:
\bigcup_{h=1}^k I_h =\cN\,,\:\ul{I}\,{\normalfont\textrm{ is connected }}
\bigg\}
\end{equation}
where $\ul{I}$ connected means that for each pair $I,I'$ in
$\ul{I}$ there exists a sequence $J_q\in \ul{I}$, $q=0,\dots,m$,  such that
$I=J_0$, $J_m=I'$, and $J_{q-1}\cap J_q\neq\emptyset$, $q=1,\dots,m$. 
We note that for $k>2^n$ we have $\cI_k=\es$.  

\begin{lem}
\label{t:srpf}
If $\dis(\Lambda_i,\Lambda_j)>r$ for any
$i\neq j\in\cN$ then for each $\tau\in\cS_{\L^\complement}$ we have 
\begin{equation}
\label{srpf0} 
\mu_\Lambda^\tau\big(f_1;\cdots;f_n\big) 
=\sum_{\sigma\in\cS_{\Lambda_{\cN}}} 
\prod_{i\in N} 
\big[\mu_{\Lambda;\Lambda_i}^\tau(\sigma_{\Lambda_i})
     f_i(\sigma_{\Lambda_i})\big]
\: \varrho^{(\Lambda)}(\sigma\tau)
\end{equation}
In particular
\begin{equation}
\label{srpf} 
\Big| \mu_\Lambda^\tau \big(f_1;\cdots;f_n\big) \Big| 
\le 
\big\| \varrho^{(\L)} \big\|_\infty\,
\prod_{i=1}^n 
\mu_\Lambda^\tau\big(|f_i|\big) 
\end{equation}
\end{lem}

\noindent{\it Proof.\/}\
For $\L\in\bF$, $F\in\cF_{\Lambda_F}$, with $\Lambda_F\subset\Lambda$,
and $\tau\in\cS_{\Lambda^\complement}$, 
by using the definition \eqref{ls-sone} of the Gibbs state, we get  
$$
\mu_\Lambda^\tau(F)= 
\sum_{\sigma\in\cS_{\Lambda_F}} 
 \frac{Z_{\Lambda\setminus\Lambda_F}(\sigma\tau)}
      {Z_{\Lambda}(\tau)}\, 
   \exp\bigg\{\sum_{\newatop{X\cap\Lambda_F\neq\emptyset}
                   {X\cap\Lambda\subset\Lambda_F}}
      U_X(\sigma\tau)\bigg\}\,
   F(\sigma) 
$$
By using (\ref{sm}), the hypotheses $\dis(\Lambda_i,\Lambda_j)>r$, 
and that $\ul{D}$ is a partition of $\cN$, we thus find 
\begin{equation}
\label{abacab}
\begin{array}{rl}
{\displaystyle
\mu_\Lambda^\tau\big(f_1;\cdots;f_n\big)=} &
{\displaystyle
\vphantom{\Bigg\{}
\sum_{\sigma\in\cS_{\L_{\cN}}}
\prod_{i=1}^n 
\bigg[
\frac{Z_{V_{\{i\}}}(\sigma\tau)}
     {Z_{\Lambda}(\tau)} 
   \exp\bigg\{\sum_{\newatop{X\cap \L_i\neq\es}
                    {X\cap\Lambda\subset\Lambda_i}}
      U_X(\sigma\tau)\bigg\}\,
   f_i(\sigma)\bigg]} \\
&
{\displaystyle\vphantom{\Bigg\{}
\;\;\;\;\times
\sum_{\ell=1}^n (-1)^{\ell -1} (\ell-1)! 
\sum_{\underline{D}\in\cD_N^\ell}
\prod_{k=1}^\ell 
\frac{Z_{V_{D_k}}(\sigma\tau)\big[Z_\Lambda(\tau)\big]^{|D_k|-1}}
     {\prod_{i\in D_k}Z_{V_{\{i\}}}(\sigma\tau)}}
\end{array}
\end{equation}

We therefore need to show that
\be{r=}
\varrho^{(\L)} = 
\sum_{\ell=1}^n (-1)^{\ell -1} (\ell-1)! 
\sum_{\underline{D}\in\cD_N^\ell}
\prod_{k=1}^\ell 
\frac{Z_{V_{D_k}} Z_\Lambda^{|D_k|-1}}
     {\prod_{i\in D_k}Z_{V_{\{i\}}} }
\end{equation}
Let $I\subset\cN$, $|I|\ge 2$, by (\ref{2.3}) we have the
following chain of identities
\begin{equation}
\label{ground}
\begin{array}{rcl}
{\displaystyle
\sum_{J\subset I}\log R_J}
&=&
{\displaystyle\vphantom{\Bigg\{_{M_{M_M}}}
\sum_{\newatop{J\subset I}{|J|\ge 2}}  
\sum_{K\subset J}  (-1)^{|J|-|K|} \log Z_{V_K}
=\sum_{K\subset I} \sum_{\newatop{J:\,I\supset J\supset K}{|J|\ge 2}}  
(-1)^{|J|-|K|} \log Z_{V_K}
}\\
&=&
{\displaystyle\vphantom{\Bigg\{^{M^{M^{M}}}}
\sum_{k=0}^{|I|}\sum_{\newatop{K\subset I}{|K|=k}}\,
\log Z_{V_K}
\sum_{j=2\vee k}^{|I|}
(-1)^{j-k}
\genfrac{(}{)}{0pt}{}{|I|-k}{j-k} 
}\\
&=&
{\displaystyle \vphantom{\bigg\{}
\log Z_{V_{I}} + (|I|-1) \log Z_\Lambda - \sum_{i\in I} \log Z_{V_{\{i\}}}}\\
\end{array}
\end{equation}
Therefore, given $\underline{D}\in\cD_N^\ell$ and $k\in\{1,\dots,\ell\}$ 
we have
\begin{equation}
\label{pip}
\prod_{J\subset D_k}(1+\varrho_J)=
\prod_{J\subset D_k} R_J=
\frac{Z_{V_{D_k}}\,Z_\Lambda^{|D_k|-1}}
     {\prod_{i\in D_k}Z_{V_{\{i\}}}}
\end{equation}
Hence, formula \eqref{r=} follows from (\ref{pip}) and 
the following identity
\begin{equation}
\label{2.5}
\varrho^{(\L)}=
\sum_{\ell=1}^n(-1)^{\ell -1} (\ell-1)! 
\sum_{\underline{D}\in\cD_N^\ell}\,
\prod_{k=1}^\ell\,
\prod_{J\subset D_k} (1+\varrho_J)
\end{equation}
where $\varrho_\emptyset=0$ (we also have $\varrho_{\{i\}}=0$ but 
this will not be used in the proof of (\ref{2.5})).

To prove (\ref{2.5}), we define
$$
{\tilde \cI}_k := \Big\{ \ul{I} \equiv \big\{I_1,\dots,I_k\big\}\,:\;
I_h\subset\cN \,,\:  h\neq h' \Rightarrow I_h\neq I_{h'}
\Big\}
$$
by expanding the products on the right hand side of \eqref{2.5} we get that 
it is equal to 
\begin{equation}\label{2.6}
\sum_{\ell=1}^n(-1)^{\ell-1}(\ell-1)!\big|\cD_N^\ell\big|+
\sum_{k\ge 1} 
\sum_{\ul{I}\in {\tilde \cI}_k} 
 a(\ul{I})\,
\prod_{I\in\ul{I}} \varrho_I 
\end{equation}
for appropriate coefficients $a(\ul{I})$ which can be computed as
follows. Let $d_n^\ell := \big|\cD_N^\ell\big|$ be the number of partitions
into $\ell$ atoms of $\cN$; we understand that $d_n^\ell=0$
if $\ell \ge n+1$.
Given $\ul{I}\in\tilde{\cI}_k$ let us decompose it into
maximal connected components $\cC_1,\dots,\cC_h$ namely,
$\ul{I}=\bigcup_{m=1}^h \cC_m$ where each $\cC_m$ is connected and for
any pair $I\in \cC_m$, $J\in \cC_{m'}$ with $m\neq m'$ we have $I\cap
J=\emptyset$; 
let also ${\tilde \cC}_m :=\bigcup_{I\in \cC_m}I\subset\cN$
and $c_m := \big| {\tilde \cC}_m  \big|$. Then 
$$
\begin{array}{rcl}
{\displaystyle  
a(\ul{I})} &=&
{\displaystyle a(\cC_1,\dots,\cC_h)
}\\
&=&
{\displaystyle 
\vphantom{\bigg\{}
 \sum_{\ell=1}^n (-1)^{\ell -1} (\ell-1)!\,
\bigg|\Big\{\underline{D}\in\cD_N^\ell \,:\; \forall m
=1,\dots,h \;\; \exists j\,\in \{1,\dots,\ell\}:\,
D_j \supset {\tilde \cC}_m \big\} \bigg|
}\\
&=&
{\displaystyle 
\vphantom{\bigg\{}
\sum_{\ell=1}^n (-1)^{\ell -1} (\ell-1)!\,  d^\ell_{n+h-(c_1+\cdots+c_h)}}
\end{array}
$$
We note that the recursion relation $d_1^1=1$ and 
$d_{i+1}^\ell=d_i^{\ell-1}+\ell\,d_i^\ell$, with $i,\ell=1,2,\dots$, holds.
Such relation implies that the first term in (\ref{2.6}) vanishes 
(recall that $n\ge 2$). Moreover, for the same reason, we
have that $a(\ul{I})=1$ if $h=1$ and $c_1=n$ namely, if
$\ul{I}\in\cI_k$, and $a(\underline{I})=0$ otherwise.  
Recalling (\ref{2.3.1}), we have thus completed 
the proof of (\ref{2.5}).
\qed

\medskip
The next Lemma states that each $\varrho_I$ has an exponential decay with the 
tree intersecting each $\Lambda_i$, $i\in I$.
Given $I=\{h_1,\dots,h_{|I|}\}\su\cN$, we define 
\begin{equation}
\label{ti}
T(I):=\inf\Big\{\bT\big(\{x_1,\dots,x_{|I|}\}\big),\,
                x_j\in\Lambda_{h_j}\textrm{ with } j=1,\dots,|I|\Big\}
\end{equation}
and note that $T(N)\ge\cT(f_1;\dots;f_n)$.

\begin{lem}
\label{t:stima1}
Let $L\in\bN$, 
assume that for each $\Lambda\in\bF_L$ and $\tau\in\cS$
we have the expansion (\ref{tm1}) as in Theorem~\ref{t:princ}
and the bound \eqref{tm2} holds.
Set $\theta_\es:=\theta_{\{i\}}:=0$ and 
\begin{equation}
\label{tri}
\theta_I:=
C\: 2^{|I|}  \: \inf_{i\in I}|\Lambda_i|
\: \exp\bigg\{- \Big[a+\frac{b}{|I|-1}\Big] T(I) \bigg\}
\end{equation}
for $|I|\ge 2$. Then, recalling $\varrho_I$ has been defined below
(\ref{2.3}), for any $I\subset\cN$ we have
\begin{equation}
\label{stima2}
\big\|\varrho_I\big\|_\infty \le
\theta_I\; e^{\theta_I}
\end{equation}
\end{lem}

\noindent{\it Proof.\/}\
By plugging (\ref{tm1}) into definition (\ref{2.3}) 
and understanding $\phi_{X,V}=0$ whenever $X\cap V=\emptyset$,
we get 
$$
\begin{array}{rcl}
\log R_I
&=& {\displaystyle 
      (-1)^{|I|}
      \sum_{j=0}^{|I|} (-1)^j
      \sum_{\newatop{J\subset I}
                    {|J|=j}} 
      \sum_{\newatop{X\subset\subset\bL}
           {X\cap V_J\not=\emptyset}}
            \phi_{X,V_J}}
= {\displaystyle 
      (-1)^{|I|}
      \sum_{j=0}^{|I|} (-1)^j
      \sum_{\newatop{X\subset\subset\bL:}
           {X\cap\Lambda\not=\emptyset}}
      \sum_{\newatop{J\subset I}
                    {|J|=j}} 
            \phi_{X,V_J}}\\
&=& {\displaystyle 
      (-1)^{|I|}
      \sum_{K\subset I}
      \;\sum_{\newatop{X\cap\Lambda\not=\emptyset,
                     \,X\cap\Lambda_{I\setminus K}=\emptyset}
                    {X\cap\Lambda_k\not=\emptyset\,\forall k\in K}}
      \;\sum_{j=0}^{|I|} (-1)^j
      \sum_{\newatop{J\subset I}
                    {|J|=j}} 
            \phi_{X,V_J}}\\
&=& {\displaystyle 
      (-1)^{|I|}
      \sum_{K\subset I}
      \;\sum_{\newatop{X\cap\Lambda\not=\emptyset,
                     \,X\cap\Lambda_{I\setminus K}=\emptyset}
                    {X\cap\Lambda_k\not=\emptyset\,\forall k\in K}}
      \;\sum_{j=0}^{|I|} (-1)^j
      \sum_{H\subset K}
      \sum_{\newatop{J\subset I}
                    {|J|=j,\,J\cap K=H}}
            \phi_{X,V_J}}\\
\end{array}
$$
We now note that the hypotheses on $\phi_{X,V}$ imply that 
$\phi_{X,V}=\phi_{X,V'}$ if 
$X\cap V^\complement=X\cap(V')^\complement$. Hence
$$
\begin{array}{rcl}
{\displaystyle \log R_I}
&=& {\displaystyle
      (-1)^{|I|}
      \sum_{K\subset I}
      \; \sum_{\newatop{X\cap\Lambda\not=\emptyset,
                     \,X\cap\Lambda_{I\setminus K}=\emptyset}
                    {X\cap\Lambda_k\not=\emptyset\,\forall k\in K}}
      \; \sum_{H\subset K}
            \phi_{X,V_H}
      \sum_{j=0}^{|I|} (-1)^j
      \sum_{\newatop{J\subset I:}
                    {|J|=j,\,J\cap K=H}} 
            1}\\
&=& {\displaystyle 
      (-1)^{|I|}
      \sum_{K\subset I}
      \; \sum_{\newatop{X\cap\Lambda\not=\emptyset,
                     \,X\cap\Lambda_{I\setminus K}=\emptyset}
                    {X\cap\Lambda_k\not=\emptyset\,\forall k\in K}}
      \; \sum_{H\subset K}
            \phi_{X,V_H}
      \sum_{j=|H|}^{|I|-|K|+|H|} (-1)^j
      \genfrac{(}{)}{0pt}{}{|I|-|K|}{j-|H|}
            }\\
&=& {\displaystyle \vphantom{\bigg\{}
      \sum_{\newatop{X\cap\Lambda\neq\emptyset}
                    {X\cap\Lambda_i\not=\emptyset\,\forall i\in I}}
      \sum_{H\subset I} (-1)^{|I|-|H|}
            \phi_{X,V_H}
            }\\
\end{array}
$$
where we used that the sum on $j$ on the second line equals
$(-1+1)^{|I|-|K|}=0$ for $K\subset I$ and $K\not=I$. 

Now, by using the bound (\ref{tm2}) and the remark that if 
$X\cap\Lambda_i\not=\emptyset$ for all $i\in I$ we have
$T(I)\le (|I|-1)\, \diam(X)$ and $T(I)\le \bT(X)$, we get
\begin{equation}
\label{arma}
\big\|\log R_I\big\|_\infty
\le
C\: 2^{|I|} \:
\inf_{i\in I}|\Lambda_i| \:
\exp\bigg\{- \Big(a+\frac{b}{|I|-1}\Big)T(I) \bigg\}
\end{equation}
which, by using the inequality $|e^u-1|\le e^{|u|}\,|u|$, implies the
bound (\ref{stima2}). 
\qed

We remark that it is not difficult to check that Lemma~\ref{t:stima1} holds 
also under the condition in Remark~\ref{3}.

\smallskip
We can now prove the first part of Theorem~\ref{t:princ}.

\smallskip
\noi{\it Proof of the bound \eqref{tm80}.\/}\
Recalling $\cI_k$ has been defined in (\ref{2.4}), it is easy to show
that for each $k\ge 1$ and $\ul{I}\in\cI_k$ we have
\be{I>f}
\sum_{I\in\ul{I}} \Big(a+\frac{b}{|I|-1}\Big)\, T(I) 
\ge
\Big(a+\frac{b}{n-1}\Big)\,\cT(f_1;\dots;f_n) 
\end{equation}
Hence \eqref{tm80} follows from Lemmata~\ref{t:srpf} and
\ref{t:stima1} provided we define $K_n$ as
\be{Kn=}
K_n(C;|\L_1|,\dots,|\L_n|) :=
\sum_{k\ge 1}\,
 \sum_{\ul{I}\in \cI_k}\,
  \prod_{I\in\ul{I}}\bigg[
C\, 2^{|I|}  \, \inf_{i\in I}|\Lambda_i| \,
\exp\big\{ C\, 2^{|I|}  \, \inf_{i\in I}|\Lambda_i|\big\}
\bigg]
\end{equation}
which is finite since it is the sum of a finite number of terms.\qed

\medskip
The constant $K_n$ in \eqref{Kn=} is highly non--optimal.
If $a>0$ we can improve the estimates and get (\ref{tm8}).

\bpro{t:sst}
Assume condition \eqref{tm6} holds for some $a>0$, $\d\in(0,1)$. 
Recalling $\theta_I$ has been defined in \eqref{tri}, set
\be{bt}
\bar\theta_I :=\exp\big\{a\,(1-\d)\,T(I) \big\} \, \theta_I
\end{equation}
then, recalling $\cI_k$ has been defined in (\ref{2.4}),
\begin{equation}
\label{trim1}
\sum_{k\ge 1}\,
 \sum_{\ul{I}\in \cI_k}\,
  \prod_{I\in\ul{I}}\,
   \bar\theta_I\,e^{\theta_I} 
\le 1
\end{equation}
\epro
To prove Proposition~\ref{t:sst} we start by a general result 
on trees which gives a lower bound on the number of edges
in term of a path connecting all the vertices. 

\blem{t:infpi}
Let $I\subset\cN$ with $2\le|I|=k+1$ be given by $I=\{i_0,i_1,\dots,i_k\}$.
Let us denote by $\Pi_0(k)$ the set of permutations $\pi$ of
$\{0,1,\dots,k\}$ such that $\pi(0)=0$. 
Recalling $T(I)$ has been defined in \eqref{ti}, we then have
\begin{equation}
\label{infpi0}
T(I) \ge 
  \frac{1}{2}
  \inf_{\pi\in\Pi_0(k)}
  \sum_{l=1}^{k} \dis\big(\L_{i_{\pi(l-1)}},\L_{i_{\pi(l)}}\big)
\end{equation}
\elem

\noi{\it Proof.\/}\
Let $X=\{\bar x_{i_0},\bar x_{i_1}, \dots,\bar x_{i_k}\}$, 
$\bar x_{i_h}\in\L_{i_h}$ be a minimizer for \eqref{ti}
and, for such $X$, let $T_X=(V_X,E_X)\subset(\bL,\bE)$ with
$V_X\supset X$ be a tree in which the infimum in \eqref{treedec0}
is attained.
The Lemma is implied by 
\begin{equation}
\label{alb}
|E_X|\ge 
  \frac{1}{2}
  \inf_{\pi\in\Pi_0(k)}
  \sum_{l=1}^{k} \dis\big(\bar x_{i_{\pi(l-1)}},\bar x_{i_{\pi(l)}}\big)
\end{equation}
which is proven as follows. By induction on the number of edges in $T_X$ 
it is easy to prove that there exists a path (see Fig.\ \ref{f:alb}) 
$\{\ell_0,\dots,\ell_{M-1}\}$, with $\ell_m\in E_X$ for all $m=0,\dots,M-1$,
satisfying the following properties:
$\ell_{m-1}\cap\ell_m\not=\emptyset$ for all $m=1,\dots,M-1$,
$\bar x_{i_0}\in\ell_0$, for each $v\in V_X$ there exists
$m\in\{0,\dots,M-1\}$ such that
$\ell_m\ni v$, and each edge $e\in E_X$ appears in the path at most twice.
Recalling that $\dis(x,y)=\tree(\{x,y\})$, the bound (\ref{alb}) follows.
\qed

\newpage
\setlength{\unitlength}{1.3pt}
\begin{figure}
\begin{picture}(200,100)(-120,-50)
\thinlines
\put(-20,20){\circle*{3}}
\put(0,40){\circle*{3}}
\put(0,0){\circle*{3}}
\put(20,60){\circle*{3}}
\put(20,20){\circle*{3}}
\put(40,40){\circle*{3}}
\put(40,0){\circle*{3}}
\put(100,40){\circle*{3}}
\put(120,60){\circle*{3}}
\put(120,20){\circle*{3}}
\put(-20,20){\line(1,1){20}}
\put(-20,20){\line(1,-1){20}}
\put(0,40){\line(1,1){20}}
\put(0,40){\line(1,-1){20}}
\put(20,20){\line(1,-1){20}}
\put(20,60){\line(1,-1){20}}
\put(100,40){\line(1,1){20}}
\put(100,40){\line(1,-1){20}}
\thinlines
\qbezier(-20,20)(-20,0)(0,0)       \put(-15,5){\vector(1,-1){1}}
\qbezier(-20,20)(0,20)(0,0)        \put(-5,15){\vector(-1,1){1}}
\qbezier(-20,20)(-20,40)(0,40)     \put(-15,35){\vector(1,1){1}}
\qbezier(0,40)(0,20)(20,20)        \put(5,25){\vector(1,-1){1}}
\qbezier(20,20)(20,0)(40,0)        \put(25,5){\vector(1,-1){1}}
\qbezier(40,0)(40,20)(20,20)       \put(35,15){\vector(-1,1){1}}
\qbezier(20,20)(20,40)(0,40)       \put(15,35){\vector(-1,1){1}}
\qbezier(0,40)(0,60)(20,60)        \put(5,55){\vector(1,1){1}}
\qbezier(20,60)(40,60)(40,40)      \put(35,55){\vector(1,-1){1}}
\qbezier[10](40,40)(40,50)(50,50)  
\qbezier[20](50,40)(70,40)(90,40)  
\qbezier[10](90,50)(100,50)(100,40)
\qbezier(100,40)(100,20)(120,20)   \put(105,25){\vector(1,-1){1}}
\qbezier(120,20)(120,40)(100,40)   \put(115,35){\vector(-1,1){1}}
\qbezier(100,40)(100,60)(120,60)   \put(105,55){\vector(1,1){1}}
\put(-32,20){$\!\bar x_{i_0}$}
\put(-22,1){$\ell_0$}
\put(0,12){$\ell_1$}
\put(-22,37){$\ell_2$}
\put(-2,21){$\ell_3$}
\put(18,1){$\ell_4$}
\put(40,12){$\ell_5$}
\put(20,32){$\ell_6$}
\put(-2,57){$\ell_7$}
\put(40,52){$\ell_8$}
\put(52,46){$\ell_9$}
\put(75,46){$\ell_{M-4}$}
\put(98,16){$\ell_{M-3}$}
\put(120,32){$\ell_{M-2}$}
\put(97,62){$\ell_{M-1}$}
\end{picture}
\vskip -1.5 cm
\caption{The path $\ell=\{\ell_0,\dots,\ell_{M-1}\}$ introduced
in the proof of Lemma~\ref{t:infpi}}
\label{f:alb}
\end{figure}
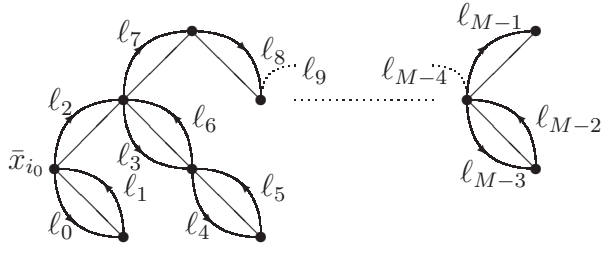

\blem{t:prep}
Assume condition \eqref{tm6} is satisfied then, recalling $\bar\theta_I$ has
been defined in \eqref{bt},
\begin{equation}\label{prep} 
{\tilde K}:= \sup_{i\in\cN}
\sum_{I\su\cN,\,I\ni i} 
(3\,e)^{|I|} \, \bar\theta_I \le \frac 13
\end{equation}
\elem

\noi{\it Proof.\/}\
By using Lemma \ref{t:infpi} and recalling that $\theta_{\{i\}}=0$ for 
all $i\in N$, we get
$$
\begin{array}{l}
{\displaystyle
\sup_{i\in\cN}
\sum_{I\su\cN,\,I\ni i} 
(3\,e)^{|I|} \,\bar\theta_I 
\;\le\; 6\,C\,e\,
\sup_{i\in\cN}\sum_{k\ge1}
\sum_{\newatop{I\su \cN\sm\{i\}}{|I|=k}} 
(6\,e)^k \Big(\inf_{j\in I}|\Lambda_j|\Big) \: e^{-a\,\d\,T(I)} 
}
\\
\qquad
{\displaystyle
\le \;6\,C\,e\,
\sup_{i_0\in\cN}\sum_{k\ge 1}
\frac{(6\,e)^k}{k!}
}
\\
\qquad
{\displaystyle
\phantom{ \le \;6Ce
}\times\;
\sum_{\newatop{i_1,\dots,i_k\in\cN\sm\{i_0\}}
	{i_h\neq i_{h'}\,,\:h\neq h'}} 
\Big(\inf_{h=0,\dots,k}|\Lambda_{i_h}|\Big) \: 
\exp\Big\{- a\,\d\,
 \frac{1}{2}\,
  \inf_{\pi\in\Pi_0(k)}
 \sum_{l=1}^{k} \dis\big(\L_{i_{\pi(l-1)}},\L_{i_{\pi(l)}}\big)\Big\}
}
\\
\qquad
{\displaystyle
\le\; 6\,C\,e\,
\sup_{i_0\in\cN}\sum_{k\ge 1}
\frac{(6\,e)^k}{k!}
\sum_{\newatop{i_1,\dots,i_k\in\cN\sm\{i_0\}}
	{i_h\neq i_{h'}\,,\:h\neq h'}} 
\sum_{\pi\in\Pi_0(k)}
\Big(\inf_{h=0,\dots,k}|\Lambda_{i_h}|\Big)
\prod_{l=1}^k 
e^{- a\,\d\,
 \frac{1}{2}\,
\dis(\L_{i_{\pi(l-1)}},\L_{i_{\pi(l)}})}
}
\\
\qquad
{\displaystyle
\le\; 6\,C\,e\,
\sum_{k=1}^\infty
(6\,e)^k \bigg(\sup_{i\in I} \sum_{j\in I\sm \{i\}}
|\L_i|\wedge |\L_j| \:
e^{- a\,\d\,
 \frac{1}{2}\, \dis(\L_j,\L_i)}
\bigg)^k
}
\\
\qquad
{\displaystyle 
\le\; 
\vphantom{\bigg\{^{M}}
6\,C\,e \,\frac {6\,e\,\l}{1-6\,e\,\l} \;\le\; \frac 13
}
\end{array}
$$
where, in the last line, we used \eqref{tm6}.
\qed

\noi{\it Proof of Proposition \ref{t:sst}.\/}\
We note that from the bound \eqref{prep} it follows
$e^{\theta_I}\le e^{\bar\theta_I}\le e$.  By letting $\e:= 1/3$ and ${\tilde
\theta}_I:= (3e)^{|I|}\bar\theta_I$ and using Lemma \ref{t:prep}, we can
apply the estimate in \cite[Appendix B]{[CasO]}
and get
$$
\sum_{k\ge 1}\, 
\sum_{\ul{I}\in \cI_k}\, 
\prod_{I\in\ul{I}}\, \bar\theta_I\,e^{\theta_I} 
\le
\sum_{k\ge 1}\, \sum_{\ul{I}\in \cI_k}\,
\prod_{I\in\ul{I}}\e^{|I|} \, {\tilde \theta}_I 
\le
\varepsilon{\tilde K} 
\left[1+\frac{e^{{\tilde K}}-1}{1
+\varepsilon^2e^{{\tilde K}} -2\varepsilon e^{{\tilde K}}}\right] 
\;\le\; 1
$$
since $\e\le1/3$ and $\tilde K \le1/3$.\qed

\smallskip
It is now straightforward to conclude the proof of Theorem~\ref{t:princ}.

\smallskip
\noi{\it Proof of the bound \eqref{tm8}.\/}\
For $\ul{I}\in\cI_k$ we have, recalling \eqref{I>f}
$$
\prod_{I\in\ul{I}} \theta_I \,e^{\theta_I}
=
\prod_{I\in\ul{I}} e^{-a\,(1-\d)\,T(I)} \,\bar\theta_I \,e^{\theta_I}
\le e^{-a\,(1-\d)\,\cT(f_1;\dots;f_n)} \, 
\prod_{I\in\ul{I}} \bar\theta_I \,e^{\theta_I}
$$
and the bound \eqref{tm8} follows from Lemmata~\ref{t:srpf},
\ref{t:stima1} and Proposition~\ref{t:sst}.\qed

\bigskip\noindent
\normalsize\textbf{Acknowledgments}
\par\noindent
{\small 
 It is a pleasure to express our thanks to B.\ Scoppola 
 for useful discussions and comments.
}

\normalfont 

 
 
 
\end{document}